\documentclass[twocolumn,aps,prd,superscriptaddress,longbibliography]{revtex4-2}

\usepackage{graphicx}
\usepackage{lmodern}
\usepackage[T1]{fontenc}
\usepackage{txfonts}
\usepackage{color}
\usepackage{xcolor}
\usepackage{hyperref}

\newcommand{\nn}{\nonumber}

\tabularnewline

\def \be {\begin{equation}}
\def \ee {\end{equation}}
\def \bea {\begin{eqnarray}}
\def \eea {\end{eqnarray}}
\newcommand{\eq}[1]{(\ref{#1})}

\begin{document}

\title{Violation of Weak Cosmic Censorship in de Sitter Space}

\author{Feng-Li Lin}
\email{fengli.lin@gmail.com}
\affiliation{Department of Physics, National Taiwan Normal University, Taipei, Taiwan 116}
\author{Bo Ning}
\email{ningbo@scu.edu.cn}
\affiliation{College of Physics, Sichuan University, Chengdu, Sichuan 610064, China}
\affiliation{Peng Huanwu Center for Fundamental Theory, Hefei, Anhui 230026, China}

\date{\today}
\begin{abstract}
Inspired by the recent discovery of a violation of strong cosmic censorship (SCC) for the near-extremal Reissner-Nordstr\"om black holes in de Sitter space (RN-dS), we investigate if the weak cosmic censorship conjecture (WCCC) can also be violated in RN-dS with a fixed cosmological constant. Our method is based on the recent formulation of examining WCCC by the constraint of the second law, which in this case requires the sum of areas of the event and cosmic horizons cannot decrease during the infall process of Wald's gedanken experiment. We find that the WCCC can be violated for the near-extremal RN-dS in some regimes of second-order perturbation of field configurations. Given the charge parameter of RN-dS, we can find the lowest value of the sub-extremality parameter, beyond which the WCCC holds. Our results imply that violations of SCC and WCCC could be correlated. Because of a lack of an unambiguous relation between the gravitational mass and matter's kinematic mass in asymptotically de Sitter space, we cannot compare the corresponding regimes of parameter space for the violations of SCC and WCCC. We also discuss the subtlety in formulating both the first-law and second-law approaches to examine WCCC. 
 \end{abstract}

\maketitle

\section{Introduction}
Curvature singularity is predicted to be unavoidable in Einstein's gravity \cite{PhysRevLett.14.57}. However, the singularity will yield unphysical results and should be shielded behind the event horizon \cite{1999JApA...20..233P}. This was formulated as the weak cosmic censorship conjecture (WCCC) \cite{Penrose:1969pc}. The WCCC can be subjected to the test of gedanken experiments \cite{Wald:1974hkz, Sorce:2017dst} by trying to destroy the event horizon of a (near-)extremal black hole via overcharging or overspinning it with some infalling matter. 
Besides, the presence of the time-like curvature singularity, shielding by an inner Cauchy horizon for the charged or rotating black holes, implies the breakdown of the determinism of Einstein's gravity. This is because the initial data cannot uniquely determine the evolution of the region beyond the Cauchy horizon. This leads to the strong cosmic censorship (SCC) \cite{1974IAUS...64...82P, Simpson:1973ua}  that the inner Cauchy horizon is unstable due to its infinite blue-shifted effect to become a null-like curvature singularity so that the determinism is preserved.

For a long time, these two forms of cosmic censorship conjectures have been considered independent. This is because they have pretty different mathematical formulations. Despite that, it is interesting to see if they imply each other, that is, if they can be satisfied or violated simultaneously. It is easier to find examples of both conjectures being satisfied, such as the Kerr-Newman black holes. The SCC is satisfied due to the mass inflation 
\cite{HISCOCK1981110, PhysRevLett.63.1663, PhysRevD.41.1796, PhysRevLett.67.789, PhysRevLett.68.2117, Brady:1995ni, Burko:1997zy} induced by the infalling perturbation as the weak solution of Einstein equation \cite{Christodoulou:2008nj}; and the WCCC is ensured by the first law of black hole mechanics and energy condition of the infalling matter \cite{Sorce:2017dst}, or by the second law of non-decreasing black hole entropy \cite{Chen:2020hjm, Lin:2022ndf}. On the other hand, the finite causal past bounded by the cosmic horizon could prevent the infinite blueshift on the Cauchy horizon so that the mass function is bounded, and the SCC could be violated for some charged black holes in de Sitter space. The study of this issue has a long history \cite{Mellor:1989ac, Brady:1992cz, Brady:1992pc, Brady:1996za, Brady:1998au}, e.g., see a review for earlier efforts in \cite{Chambers:1997ef}. Inspired by the mathematical study of Cauchy evolution in the black hole background as the weak solutions \cite{Dafermos:2003wr, Dafermos:2012np, Costa:2014yha, Hintz:2015jkj, Costa:2016afl, Cardoso:2017soq, dafermos2017interior, Costa:2017tjc, luk2017strong, luk2019strong}, a revival of interest in this issue was initiated in \cite{Cardoso:2017soq, Cardoso:2018nvb} and followed up by \cite{Dias:2018ufh}. The criterion of examining SCC is mainly done by reformulating the SCC as the inextendibility of the maximal Cauchy evolution as a weak solution of the field equations with smooth initial data so that the stability of the Cauchy horizon is related to the spectral gap of the quasi-normal modes (QNM) of the background spacetime induced by infalling matter \cite{Hintz:2015jkj, 2016arXiv161204489H, Costa:2016afl, Costa:2017tjc}. Thus, if the spectral gap $\beta:=- \rm{Im}(\omega_0)/\kappa_-$ is less than $1/2$, where $\omega_0$ is the frequency of the longest-lived QNM and $\kappa_-$ the surface gravity of Cauchy horizon, then SCC holds. In \cite{Cardoso:2017soq, Cardoso:2018nvb, Dias:2018ufh}, by numerically solving the QNM of (charged) scalar waves, the parameter ranges of violating SCC for the near-extremal black holes are found.

As far as we know, the WCCC for the Reissner–Nordström (RN) black holes in de Siter space has not yet been examined. Naively, we may generalize Wald's gedanken experiments to examine WCCC by trying to overspin or overcharge a (near-)extremal black hole with infalling matter \cite{Wald:1974hkz, Sorce:2017dst}. However, we will see that the existence of the cosmic horizon hinders the direct application of Wald's method based on transforming matter's energy conservation to the first law of black hole mechanics. To see such difficulty, let's recall Wald's procedure \cite{Wald:1974hkz, Chen:2020hjm} for examining the WCCC for an extremal RN black hole of mass $M$ and charge $Q=M$ in flat spacetime by throwing into it a particle of mass $m$ and charge $q$. If the particle is freely falling, its conserved energy is given by
\be
E=-(m u_{\mu} + q A_{\mu}) t^{\mu}
\ee
where $A_{\mu}$ and $t^{\mu}$ are the Maxwell gauge field and the time-like Killing vector of the background configuration. We can evaluate $E$ either on the event horizon at $r=r_+$ or at spatial infinity $r=\infty$, and yield the relation
\be\label{E_conserved}
-(m u_{\mu} + q A_{\mu}) t^{\mu}\big\vert_{r=r_+} = - m u_{\mu} t^{\mu} \big\vert_{r=\infty}
\ee
where we have used the fact that $\lim_{r\rightarrow \infty} A_{\mu}=0$.

As the particle goes into the black hole, it causes a change of the mass and charge by the amount $\delta M$ and $\delta Q$. It is known that the gravitational mass cannot be defined locally but can be defined with Arnowitt–Deser–Misner (ADM) formalism at spatial infinity, i.e., $r=\infty$. Therefore, we can then relate $\delta M$ to $E$ at spatial infinity, i.e.,
\be \label{ADM}
\delta M = - m u_{\mu} t^{\mu} \big\vert_{r=\infty}\;.
\ee
Along with the relation $\delta Q=q$ ensured by Gauss's law and the fact that $A_{\mu}t^{\mu}\big\vert_{r=r_+}=-1$ for the extremal black hole, we can convert the relation \eq{E_conserved} for the conservation of freely falling particle into the following first law of black hole mechanics, 
\be\label{1st_law} 
\delta M - \delta Q = -m u_{\mu}t^{\mu}\big\vert_{r=r_+} \ge 0\;.
\ee
The last inequality is ensured by the energy condition, i.e., $m \ge 0$, and $-u_{\mu}t^{\mu}\big\vert_{r=r_+} \ge 0$ for an ingoing particle. Therefore, we see that the WCCC is preserved because $M+\delta M\ge Q+\delta Q$.

From the above discussion, we can see \eq{ADM} relating the ADM mass $\delta M$ to $m$ is a key ingredient to arrive at the first law relation \eq{1st_law} for checking WCCC. However, the cosmic horizon prevents us from defining the ADM mass $\delta M$ properly.     Therefore, to examine the WCCC for a black hole in de Sitter space, we need a method that does not rely on the definition of ADM quantities. Indeed, in \cite{Lin:2022ndf}, an alternative method for checking WCCC based on the second law has been proposed. The second law, which requires the Wald entropy never to decrease, provides a constraint relating the changes of mass and charges of the black hole due to the infalling matter. This constraint then ensures WCCC if it holds. For the Kerr-Newman black holes, both this second-law method and the one proposed in \cite{Sorce:2017dst} based on the first law yield WCCC. The WCCC for the Myers-Perry black holes examined by the second-law method has also been verified, as shown in Appendix \ref{app_1}. To generalize this second-law method to the black holes in de Sitter space, we face the problem of defining the gravitational entropy for such background. Due to the lack of asymptotic spatial infinity to determine the ADM quantity and thus the first law, gravitational entropy is also ambiguous. There are various definitions for the gravitational entropy for black holes in de Sitter space; see \cite{Bousso:2000nf, Shankaranarayanan:2003ya, Urano:2009xn, Bhattacharya:2015mja} for examples. Among them, we take the gravitational entropy to be \cite{Cai:1997ih, Urano:2009xn, Bhattacharya:2015mja}
\be\label{S_W}
S= \pi (r_+^2 + r_c^2)
\ee
where $r_+$ and $r_c$ are the radii of event and cosmic horizons, respectively. The additive feature of \eq{S_W} agrees with the extensive nature of entropy. Moreover, one of the physical interpretations of entropy is to characterize the ignorance of information. This is consistent with the gravitational entropy given in \eq{S_W}, which characterizes the ignorance of information behind both the event and cosmic horizons for the observer in the region between these two horizons \cite{Shaghoulian:2022fop}. This contrasts the entropy bound proposed in \cite{Bousso:2000nf} for the observable entropy in de Sitter space. 
Finally, one can view \eq{S_W} as the multi-horizon generalization of Wald's covariant entropy formula \cite{Wald:1993nt, Iyer:1994ys} for the single horizon case, in which the gravitational entropy is obtained by evaluating the Noether charge on the horizon, i.e., the boundary of the Cauchy hypersurface.

For the case without a cosmic horizon, the second law requires $r_+$ always to increase. With the presence of $r_c^2$ term in $S$ of \eq{S_W} for the black hole in de Sitter space, the second law may not always guarantee the increase of $r_+$. For example, some infalling matter may cause the decrease of $r_+$ and the increase of $r_c$, such that the overall entropy change is positive. Thus, for a (near-)extremal RN black hole in de Sitter space, it is possible to violate WCCC without violating the second law.

In this paper, we will examine the above situations in detail and determine the regime in parameter space for which the second law fails to ensure WCCC in de Sitter space. Our result is in accordance with the violation of SCC. However, a direct comparison for the regime of parameter space is not viable due to the lack of direct relation between the mass/charges of the infalling matter and the resultant changes of the black hole's mass/charge in de Sitter space. This is also why one cannot adopt Wald's first law formulation to check WCCC, as discussed. Despite that, it is the first example for which one can show that both WCCC and SCC are violated.  

The remainder of the paper is organized as follows. In the next section, we solve the horizons of the RN black holes in de Sitter space (RN-dS). In section \ref{sec_3}, we recapitulate the second-law method for checking WCCC and apply it to a limiting case of RN-dS. We present our main result in section \ref{sec_4} by examining the WCCC for RN-dS with the second-law method. We then identify the WCCC-violating regime for both first- and second-order variations.  Finally, we conclude our paper in section \ref{sec_5} with some discussion on the subtlety in formulating both the first-law and second-law approaches in examining the WCCC. In Appendix A, we give the details of proving WCCC for 5-dimensional (5D) Myers-Perry black holes by both the second-law and first-law methods.

\section{RN black hole in de Sitter space}

Consider an RN  black hole with the mass parameter $M$ and charge parameter $Q$ in de Sitter space of cosmic constant $\Lambda =3 / L^2$.  We will abbreviate it as RN-dS. Then, the metric is given by
\be
ds^2 ~=~ - f(r) \, dt^2 \,+\, \frac{dr^2}{f(r)} \,+\, r^2 d\Omega_2^2 \;
\ee
and the gauge potential by
\be
A ~=~ - \, {Q \over r} \,dt \;,
\ee
in which 
\be \label{fr}
f(r) ~=~ 1 \,-\, {r^2 \over L^2} \,-\, {2 M \over r} \,+\, {Q^2 \over r^2}  \;.
\ee
In reasonable parameter range, $\,f(r) = 0\,$ has three real positive roots $\,r_- \,\le \,r_+ \, \le \, r_c \,$, corresponding to the Cauchy horizon, event horizon and cosmic horizon, respectively. The equation $\,f(r) = 0\,$ is of the form of so-called depressed quartic equation
\be \label{depqr}
r^4 \,+\, a r^2 \,+\, b r \,+\, c \;=\; 0 \,
\ee
with $\,a = -L^2\,,~ b = 2 L^2 M\,$ and $\, c = - L^2 Q^2\,$, which could be solved in terms of solutions of a related cubic equation as follows. Let $\,y\,$ be any solution of the cubic equation
\be \label{cuby}
y^2 \,-\, c \;\;=\;\; \frac{b^2}{\,4\,(2y - a)} \;\,,
\ee 
then (\ref{depqr}) could be converted to an equation with both sides perfect squares that could be solved by 
\be
r ~=~ {1 \over 2} \,\left( - \,\eta_1 \sqrt{2 y - a} \;+ \,\eta_2  \sqrt{- 2 y - a + \eta_1 \frac{2 b}{\sqrt{2 y - a}} } ~ \right) \;, 
\ee
in which $\,\eta_{1,\,2} \,=\, \pm 1\,$. The cubic equation (\ref{cuby}) can be solved in a standard procedure. Let
\bea
p &=& - \, {a^2 \over 12 } \,-\, c \;,  \\
q &=& - \, {a^3 \over 108 } \,+\, { a c \over 3} \,-\, { b^2 \over 8} \;,
\eea
(\ref{cuby}) has the solution
\be
y \;\;=\;\; {a \over 6} \,+\, w \,-\, {p \over 3 w} \;,
\ee
with
\be
w \;=\; \sqrt[3]{ - {q \over 2} + \sqrt{  {q^2 \over 4} + {p^3 \over 27 } } \, }  \;. 
\ee

For (\ref{depqr}), single double-root happens when
\be \label{sm2}
\left( a^2 \,+\, 12 c \right)^3 \;=\; \left( a\left(a^2 \,-\, 36 c\right) \,+\, {27  \over 2} b^2 \right)^2 \;>\; 0 \;\neq\; b \;\,,
\ee
which corresponds to either extremal black hole with $\,r_+ = r_-\,$ or Nariai-like spacetime with $\,r_+ = r_c\,$. The extremal condition turns out to be
\be \label{ex}
M \;\;-\;\; {1 \over 3\sqrt{6} } \,\left( L^2 \;+\; 36 Q^2 \;-\; \frac{\left( L^4 \,-\, 12 L^2 Q^2 \right)^{3/2}}{L^4} \, \right)^{1/2} =0 \;\;.
\ee
for which we have 
\bea \label{rp}
r_+ &=& r_- \;\;=\;\; \sqrt{\, {1 \over 6}\, L \left(\, L \,-\, \sqrt{L^2 \,-\, 12 Q^2 } \; \right)  }  \;\;,  \\
r_c &=& - \;\, \sqrt{\, {1 \over 6}\, L \left(\, L \,-\, \sqrt{L^2 \,-\, 12 Q^2 } \; \right)  }  \nn \\
&& \;\,+\;\, \sqrt{\, {1 \over 3}\, L \left(\,2 L \,+\, \sqrt{L^2 \,-\, 12 Q^2 } \; \right)  } \;\;.  
\eea
In terms of $\,r_+\,,\; r_c\;$, the extremal condition (\ref{ex}) could be expressed as 
\be \label{ex2}
r_c \;\;=\;\,  \sqrt{\,L^2 \,-\, 2 r_+^{2}\,} \,-\, r_+  \;.
\ee

The entropy for RN-dS, either extremal or non-extremal, contains contributions from both the event horizon and cosmic horizon obeying the area law \eq{S_W}. For simplicity, hereafter we replace $\,r/L \to r\,$, $\,M/L \to M\,$ and $\,Q/L \to Q\,$, i.e., equivalently we choose $\,L = 1\,$.

\section{WCCC and second law} \label{sec_3}

The WCCC can be examined by trying to overspin or overcharge a black hole and see if the event horizon can be destroyed. A viable way is to follow Wald's proposed gedanken experiment \cite{Wald:1974hkz} by throwing into matter obeying the energy condition into a (near-)extremal black hole. In a seminal paper \cite{Sorce:2017dst}, the WCCC has been shown to hold in Einstein-Maxwell theory by adopting Wald's covariant formulation of the first law of black hole mechanics. The first law relates the changes of mass, spin, and charges of a black hole to the corresponding quantities of the infalling matter. It ensures that the infalling matter of positive energy cannot decrease the black hole horizon. This energetic consideration for WCCC avoids the difficulty of dealing with the backreaction and self-force when directly studying the dynamical problem of the infalling matter.
Later, this energetic approach is generalized to show that WCCC also holds for the extremal non-spinning charged black holes \cite{Chen:2020hjm} in generic higher derivative gravity and electromagnetic theories.

When applying the energetic consideration of \cite{Sorce:2017dst} to the near-extremal black holes, it needs to take into account the backreaction of the second-order gravitational and electromagnetic waves. This requires finding the corresponding stress tensors in higher derivative theories. Unfortunately, the covariant construction of such stress tensors is complicated and ambiguous, e.g., \cite{Stein:2010pn}. Naively assuming no such backreaction, the energetic consideration of \cite{Sorce:2017dst} yields failure of WCCC as checked in \cite{Lin:2022ndf}. To bypass such difficulty, in \cite{Lin:2022ndf}, the second-law method, as discussed earlier, is proposed to check the WCCC for the near-extremal non-spinning charged black holes in higher derivative theories. The results show that the WCCC holds for these black holes. In Einstein's gravity, the second law can be proved equivalent to the first law and the imposition of the null energy condition of matter \cite{Hawking:1971tu, Hollands:2022fkn}. Indeed, the second law method gives the same result as the first law to ensure WCCC for the near-extremal Kerr-Newman black holes, as shown in \cite{Lin:2022ndf}.

The advantage of the second law formulation is its simplicity compared to the first law, without requiring the input of the energy condition of the infalling matter and the ADM quantities. The key idea for WCCC is to answer what kind of matter can fall into a black hole or how the conserved quantities of a black hole can vary. Thus, the spirit of the second law is that variations of these quantities should not violate the second law. For a black hole in the general theories of gravity, the entropy can be constructed by Wald's covariant Noether's charge method. Let us denote it as  
\be
S=S_{\rm Wald}(M,Q,J) 
\ee
for a black hole with mass $M$, charge $Q$ and angular momentum $J$.

The infalling matter then causes variations of the conserved quantities, i.e.,
\be
{\cal C} \rightarrow {\cal C}+ \sum_n {\lambda^n \over n!} \delta^n {\cal C}, \qquad {\cal C}=M, Q, J\;.
\ee
Here, we introduce the parameter $0\le |\lambda| <1$ to track the order of variations. The allowed variations are required to satisfy the second law order by order in $\lambda$, i.e.,
\be\label{2nd law}
\delta^n S\ge 0, \;\; {\rm for \; all}\;  n
\ee
with $\delta^n S$ defined by 
\be
\delta S=S\Big(C+\sum_n {\lambda^n \over n!} \delta^n {\cal C}\Big) - S({\cal C}) := \sum_n {\lambda^n \over n!} \delta^n S\;.
\ee
Unlike the first law, we can choose the quantities ${\cal C}$ to be the ones other than $M$, $Q$, and $J$ as long as the number of independent parameters is the same, such as the radii of black hole's and cosmic horizons or some other geometrical quantities, to proceed with the variations.  Thus, the WCCC check based on the second law can be done even for cases with no proper ADM quantities, such as the RN-dS.

The WCCC can be violated if the resultant black hole after infall becomes super-extremal. Thus, it is essential to find out the sub-extremal condition, which we abstractly denote as 
\be\label{Wsub}
W({\cal C}) \ge 0\;.
\ee
for a black hole characterized by a set of parameters denoted by ${\cal C}$. The exact form of $W({\cal C})$ depends on the types of black holes and the choices of ${\cal C}$. For RN-dS, $W(M, Q)$ is given by the LHS of \eq{ex} for a given $L$.

For WCCC to hold, we should require that the variations of $C$ satisfying the second law constraints should maintain the sub-extremal condition after the infall, this can be summarized succinctly by 
\be\label{wccc}
W_k({\cal C}+\delta {\cal C}):=W\Big({\cal C}+\sum_{n=1}^k {\lambda^n \over n!} \delta^n {\cal C}\Big\vert_{\rm 2nd \; law}\Big) \ge 0 
\ee
for all $0 \le |\lambda| <1$. In practice, we will consider the lower bound of $W_k({\cal C}+\delta {\cal C})$, denoted as $W_k(\lambda)$ and given by 
\be\label{wccc_d}
W_k(\lambda):= W\Big({\cal C}+\sum_{n=1}^k {\lambda^n \over n!} \delta^n {\cal C}\Big\vert_{\delta^1 S=\delta^2 S=\cdots=\delta^k S=0}\Big)\;.
\ee
We call $W_k(\lambda)$ the WCCC discriminant function. Note that the index $k$ indicates the maximal order of variations. In this paper, we will only consider $k=1,2$. If $W_k(\lambda)$'s are positive definite for all $k$ considered, then WCCC is ensured by the second law. Otherwise, WCCC can be possibly violated.

For the purpose of proving WCCC, we can follow the spirit of \cite{Sorce:2017dst} by first considering extremal black holes, for which there is no backreaction due to radiation flux during the infall process so that it is enough to require
$\delta^1 S \ge 0$ to prove $W_1 (\lambda)\ge 0$ for all $\lambda$. For the near-extremal black holes, the second-order backreaction effect requires checking WCCC up to $\delta^2 {\cal C}$. We can then assume $\delta^1 {\cal C}$ is optimally done to make
\be\label{n=1}
\delta^1 S =0\;,
\ee
but requires $\delta^2 C$ to obey 
\be\label{n=2}
\delta^2 S\ge 0
\ee
to check if $W_2(\lambda)$ is a complete square or not. For the Einstein-Maxwell gravity, for which the first law formulation by \cite{Sorce:2017dst} can be worked out precisely, it was shown in \cite{Lin:2022ndf} that both the first and second law formulations give the same complete square for the LHS of \eq{wccc}. Namely, for near-extremal RN black holes with $Q=M\sqrt{1-\epsilon^2}$ with $\epsilon$ parametrizing the sub-extremality and the WCCC discriminant function $W(M, Q)=M^2-Q^2 $, the conditions \eq{n=1} and \eq{n=2} give \cite{Sorce:2017dst, Lin:2022ndf}, 
\bea
&& \delta^1 S=0 \rightarrow \delta^1 M=(1-\epsilon) \delta^1 Q\;, \\
&&  \delta^2 S=0 \rightarrow \delta^2 M\ge  {1-\epsilon \over M} (\delta^1 Q)^2 +\Big(1-\epsilon +{\epsilon^2 \over 2} \Big) \delta^2Q\;, \qquad 
\eea
which then yield
\be\label{RN_W}
 W_2(\lambda)=(M \epsilon -\lambda \delta^1 Q)^2\;. 
\ee
Note that it is a complete square so that WCCC holds either sign of $\lambda$. This demonstrates the validity of the second law formulation of checking WCCC. As discussed, we can choose different ${\cal C}$ and $W({\cal C})$ to examine WCCC. For RN case, we can choose ${\cal C}=\{r_+,r_-\}$, and $W({\cal C})= r_+$ for the sub-extremal condition \eq{Wsub}. Since the Wald entropy is simply $S=\pi r_+^2$, thus the second law $\delta S= \pi r_+ \delta r_+ \ge 0$ implies $\delta r_+ \ge 0$, which then ensures WCCC. The simplicity of the above proof demonstrates the advantage of choosing proper ${\cal C}$ and $W({\cal C})$ when examining WCCC.

On the other hand, when considering the cases with possible violations of WCCC, such as the RN-dS, our purpose is to identify the regimes of parameters such as ${\cal C}$ or $\lambda$, for which the second law constraints can no longer guarantee WCCC. That is, we like to find the regimes of parameters $Q$, $\epsilon$ and $\lambda$ such that $W_k(\lambda) \le 0$ so that  \eq{wccc} cannot hold definitely. This implies that the WCCC can be possibly violated because the second law does not guarantee the validity of \eq{wccc}. Thus, we need to adopt a slightly different strategy than the one above to prove WCCC. In such cases, WCCC may be violated at the first-order variations, i.e., $W_1(\lambda)$ cannot be positive definite for $\delta^1 {\cal C}$ satisfying $\delta^1 S\ge 0$. Then, it seems there is no need to turn on the variations of second or higher orders as $W_1(\lambda)$ for the purpose of disproving WCCC. However, from the experience of proving WCCC, the second-order variations will tend to save WCCC. Therefore, to pin down a more precise WCCC-violating regime of parameter space, we should turn on $\delta^2 {\cal C}$. Again, we can assume $\delta^1 {\cal C}$ is optimally done, i.e., $\delta^1 S=0$, and then introduce $\delta^2 {\cal C}$ satisfying $\delta^2 S \ge 0$ to examine the difference between $W_1(\lambda)$ and $W_2(\lambda)$. As we will see, the latter usually yields a narrower WCCC-violating regime than the former.

\subsection{WCCC in the limiting de Sitter}

Before considering WCCC for RN-dS with a sizable cosmological constant such that $\alpha:= M/L$ is finite based on the second-law method with entropy given by \eq{S_W}, we first apply the method to check WCCC for the limiting case with an infinitesimal $\alpha$ and expand the result in the order of $\alpha$.

We start by expanding all the quantities up to ${\cal O}(\alpha^0)$, thus
\be\label{alpha0}
r_+=M+\sqrt{M^2-Q^2}\;, \quad r_c=M(-1+{1\over \alpha})\;,
\ee
and the sub-extremal condition is 
\be\label{wccc_p_alpha0}
W(M,Q)=M^2-Q^2 \ge 0\;,
\ee
which was obtained by plugging \eq{alpha0} into the extremal condition \eq{ex2} and expanding it up to ${\cal O}(\alpha^0)$.
We then consider a near-extremal black hole with 
\be
Q=M\sqrt{1-\epsilon^2} \quad {\rm with }\quad  0< \epsilon \ll 1 
\ee
where $\epsilon$ characterizes the size of sub-extremality.
Then, using the condition \eq{n=1} and \eq{n=2} up to  ${\cal O}(\alpha^0)$ for the entropy $S$ given by \eq{S_W}, we find that the resultant WCCC discriminant of \eq{wccc_d} is
\be\label{wccc_alpha0}
W_2(\lambda)=\Big( M \epsilon - (2-{2\over \alpha}+{1\over \alpha^2}) \lambda \delta^1 Q \Big)^2\;,
\ee
which is a complete square so that WCCC holds at ${\cal O}(\alpha^0)$. We see that \eq{wccc_alpha0} is singular in the limit of $\alpha \rightarrow 0$, and cannot smoothly recover \eq{RN_W} for a near-extremal RN black hole in flat spacetime. This suggests that the check for WCCC by the expansion in $\alpha$ will break down when going to the higher order of $\alpha$ expansion.

To see this, we expand all the quantities up to ${\cal O}(\alpha)$. In this case, the sub-extremal condition is the same as \eq{wccc_p_alpha0} up to an irrelevant factor $(1+4 \alpha)$, which we will omit. Following the same procedure for ${\cal O}(\alpha^0)$ case, we find that the resultant $W_2(\lambda)$ cannot be a complete square at ${\cal O}(\alpha)$ unless  ${\cal O}(\alpha^2)$ terms are added. Note that this is different from \eq{wccc_alpha0}, which is a complete square within ${\cal O}(\alpha^0)$. The need of  ${\cal O}(\alpha^2)$ terms for checking WCCC at ${\cal O}(\alpha)$ motivates to go to ${\cal O}(\alpha^2)$ or higher. However, the singular nature of $\alpha$ expansion appears at  ${\cal O}(\alpha^2)$ by having
\bea
r_+ &=& M+\sqrt{M^2-Q^2}\nn \\
&& +\; \alpha^2 \Big( {2(2 M^2-Q^2) \over M}  +{8M^4-8M^2 Q^2+Q^4 \over 2M^2 \sqrt{M^2-Q^2}}\Big)\;. \qquad 
\eea
The ${\cal O}(\alpha^2)$ term of $r_+$ is singular at the extremal limit $M^2=Q^2$, which further causes the ambiguity when solving the second law constraints \eq{n=1} and \eq{n=2}, with the solutions depending on the order of truncating the expansions of $\epsilon$ and $\alpha$ to the second order. This then fails the check of WCCC at ${\cal O}(\alpha^{m\ge 2})$.
We can conclude that the $\alpha \rightarrow 0$ is the singular limit for the RN-dS, so we cannot consistently check WCCC by the $\alpha$ expansion. In the next section, we will examine WCCC for the finite $\alpha$ case without relying on $\alpha$ expansion.

\section{Violations of WCCC for RN black holes in de Sitter} \label{sec_4}

Based on the scheme discussed in the previous section, we now examine the WCCC for near-extremal RN-dS and determine the WCCC-violating regimes of parameter space based on the second-law method with the entropy of RN-dS given by \eq{S_W}. We first consider the linear-order variations and then the second-order variations.

Instead of choosing $M$ and $Q$ to define the sub-extremal condition, we choose $r_+$ and $r_c$ to make the overall calculations less tedious because the entropy formula \eq{S_W} of RN-dS has a simple form in terms of them. From (\ref{ex2}), we can define the following sub-extremal condition for RN-dS,
\be \label{crW2}
W(r_+, \,r_c) \,=\, r_c \,-\, \left( \sqrt{1 - 2 r_+^{2}} - r_+ \right) \, \ge \, 0\;,
\ee
and introduce the sub-extremal parameter $\epsilon$ accordingly as following,
\be \label{nex2}
r_c \;=\; \left( \sqrt{1 - 2 r_+^{2}} - r_+ \right) \sqrt{1 + \epsilon}\;.
\ee
Note also that it should require $r_+$$ \le 1/\sqrt{2}$ for \eq{crW2} and \eq{nex2} to make sense.

\subsection{Linear order variation}

Consider a sub-extremal RN-dS perturbed by infalling matter. By using (\ref{nex2}), the linear order variation inequality $\,\delta^1 S(r_+,\,r_c) \,\geq\, 0\,$ gives
\be \label{varI}
\delta r_c \,\geq\, - \, \frac{( \,2 - \epsilon \,) \, r_+ } {2 \left( \sqrt{1 - 2 r_+^{2}\,} - r_+\right)} \,\delta r_+ \;.
\ee  
As $\epsilon$ is a small parameter, it is interesting to see that, in general, $\delta r_c$ has the opposite sign to $\delta r_+$ as required by the second law. That is, for the second law to hold, the cosmic and event horizons cannot be increased or decreased at the same time by the infalling matter. Especially, it is possible that the decrease of black hole entropy due to the shrinking of the event horizon can be compensated by the increase of the cosmic horizon by the expansion of the cosmic horizon to satisfy the second law. This can possibly lead to the destruction of the black hole and violation of WCCC and is incredible for asymptotic flat or AdS black holes with no cosmological horizon.

Assuming that the resultant field configuration is again an RN-dS and belongs to a one-parameter family of solutions to field equations, described by
\be
r_+(\lambda) \;=\; r_+ +  \lambda \delta r_+ \;,  \quad \quad
r_c(\lambda) \;=\; r_c +  \lambda \delta r_c  \;\;,
\ee
in which $\,\lambda\,$ is the small parameter characterizing perturbation and is assumed to be of $\,{\cal O}(\epsilon)\,$. The WCCC discriminant function $\,W_1(\lambda)\,$ for the first-order perturbation based on (\ref{crW2}) turns out to be 
\bea \label{crW2Iv}
W_1(\lambda) &=&  {1 \over 2} \left( \sqrt{1 - 2 r_+^{2}\,} - r_+ \right) \epsilon   \nn \\ 
&& +\;\, \frac{ 4 r_+^2 - 1 }{  2 r_+^{2} - 1 + r_+ \sqrt{1 - 2 r_+^{2}}  } \;\lambda\delta r_+ \;\,,
\eea
in which we have used the defining equation (\ref{nex2}) for the sub-extremality and the saturation of the second law from (\ref{varI}). Recall that by construction, any linear perturbation obeying (\ref{crW2Iv}) due to the infall process will saturate the second law. Given $\epsilon$ and $r_+$, the RHS of  (\ref{crW2Iv}) is just a linear function of the parameter $\lambda$, and cannot be positive definite. If $W_1(\lambda)<0$, it implies that the second law cannot guarantee \eq{wccc} so that WCCC can be violated by some infalling matter.

\begin{figure}[hbt!]  
  \centering
  \includegraphics[width=1\linewidth]{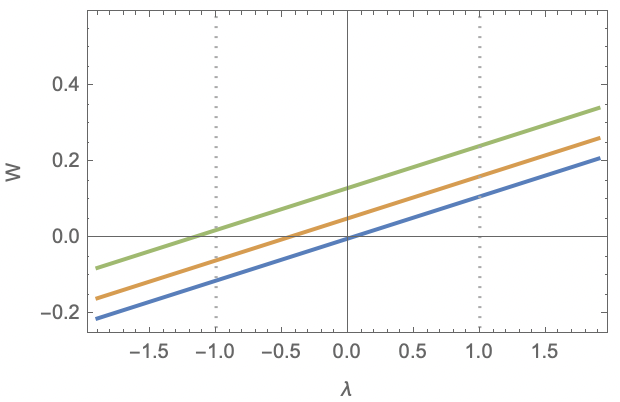}
  \caption{ The WCCC  discriminant function $W_1(\lambda)$ saturating the second law by the 
 linear order one-parameter perturbation of RN-dS configurations. We choose $\,Q = 0.1\,$ which  corresponds to extremal event horizon radius $\,r_+ = 0.102\,$. $\,\delta r_+$ and $\,\delta^2 r_+$  should be of $\,{\cal O}(r_+)\,$, hence we simply choose $\,\delta r_+ = \delta^2 r_+ = r_+\,$. The blue, orange and green lines correspond to $\,\epsilon = 0\,$, $\, 0.119965\,$ and $\,0.3\,$, respectively. }
  \label{fig:1st}
\end{figure}

The RHS of (\ref{crW2Iv}) is plotted as a function of $\,\lambda\,$ for different values of $\,\epsilon\,$ in Fig. \ref{fig:1st}. Since the extremal RN-dS black hole is characterized by only one parameter, for numerical calculation one could choose either $\,r_+\,$ or equivalently $\,Q\,$ (related by (\ref{rp})) as the free parameter. We choose the black hole charge parameter $\,Q = 0.1\,$, corresponding to extremal event horizon radius $\,r_+ = 0.102\,$. The blue line with $\,\epsilon = 0\,$ is the extremal RN-dS case, showing that as long as $\,\lambda < 0\,$, WCCC would definitely be violated. As we have discussed, this is understandable as the second law can be maintained by shrinking the event horizon but expanding the cosmic horizon. As shown, this scenario can be realized for negative $\lambda$.

For the near-extremal RN-dS, the orange and green lines in Fig. \ref{fig:1st} show that WCCC might also be violated for sufficiently large $\,|\lambda|\,$. However, since $\,|\lambda|\,$ is assumed to be small, only the $\,|\lambda| < 1\,$ regime of these lines is reliable, as is highlighted by the vertically dotted gray lines in \ref{fig:1st}. This indicates that WCCC might be restored for non-extremal RN-dS black holes. To see if the second-order perturbation could make things clearer, we need to examine $\,\delta^2 S\,$.

\subsection{Second order variation}

Following the same concern as in \cite{Sorce:2017dst, Lin:2022ndf} for examining WCCC of a near-extremal black hole, we need to consider the variation of the field configurations up to the second order to take care of the backreaction effect.  To see the effect of the second-order perturbation, we start with the optimally-done first-order perturbation by requiring $\delta^1 S = 0\,$. This gives
\be \label{1varII}
\delta r_c \;\;=\;\; -\; \frac{(\,8 \,-\, 4 \,\epsilon \,+\, 3\, \epsilon^2\,) \, r_+ } {8 \left( \sqrt{\,1 \,-\, 2\, r_+^{2}\,} \,-\, r_+\right)} \;\delta r_+ \;,
\ee
for which we have kept up to ${\cal O}(\epsilon^2)$ for consistency.

\begin{figure}[hbt!]  
  \centering
  \includegraphics[width=1\linewidth]{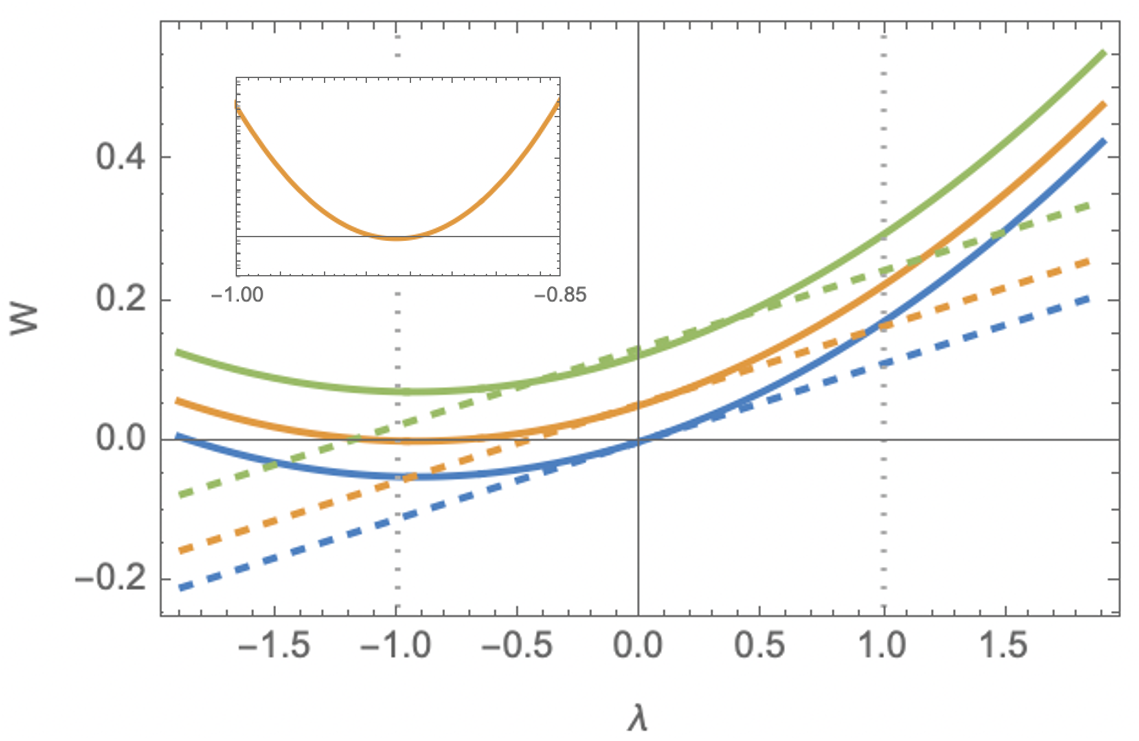}
  \caption{ The WCCC discriminant function $\, W_{2\,}(\lambda)\,$ saturating the second laws by both linear and second-order one-parameter perturbations of RN-dS configurations. The entropy is chosen to be the one given by \eq{S_W}. The results of corresponding $\,W_{1\,}(\lambda)\,$ in Fig. \ref{fig:1st} are also shown in dashed lines for comparison. We choose again $\,Q = 0.1\,$ which corresponds to extremal event horizon $\,r_+ = 0.102\,$;  and $\,\delta r_+ = \delta^2 r_+ = r_+\,$. The blue, orange and green lines again correspond to $\,\epsilon = 0\,$, $\epsilon_c$ and $\,0.3\,$, respectively.  Here, $\epsilon_c:=0.119965$ is the critical value of $\epsilon$, beyond which $W_2(\lambda)$ is positive definite. The inset figure is to show this zoomed-in region near the local minimum of the $\epsilon= \epsilon_c$ curve, at which $W_2(\lambda)=0$. Our results show that the WCCC is violated only for near-extremal RN-dS for $\epsilon < \epsilon_c$. This is qualitatively consistent with the violation of SCC.  }
  \label{fig:2nd}
\end{figure}

On top of that, we further require the second-order perturbations to satisfy the second-order second law, $\delta^2 S \ge 0\,$, and it gives 
\begin{widetext}
\bea \label{2varII}
\delta^2 r_c &\geq& \frac{  - 8 \,+\, 4 \epsilon \,+\, 4 \epsilon \left(\,2 \,-\, 3 \epsilon \,\right)r_+^2 \,-\, 3 \epsilon^2 \,+\, 2 \left(\,8 \,-\, 4 \epsilon \,+\, 3 \epsilon^2 \,\right) r_+ \sqrt{\,1 \,-\, 2\, r_+^{2}\,} }{8 \left( - \, 3 r_+ \,+\, 5 r_+^3  \,+\,  \left(\, 1 + r_+^2 \right) \sqrt{\,1 \,-\, 2\, r_+^{2}\,} \, \right)} \;(\delta r_+)^2   \nn  \\
&&~ +\;\; 
 \frac{ \left(\, 8 \,-\, 4 \epsilon \,+\, 3 \epsilon^2 \,\right) r_+ \left( - 1 \,+\, r_+^2 + 2 r_+ \sqrt{\,1 \,-\, 2\, r_+^{2}\,} \right) }{8 \left(- \, 3 r_+ \,+\, 5 r_+^3\,+\,  \left(\, 1 + r_+^2 \right) \sqrt{\,1 \,-\, 2\, r_+^{2}\,} \, \right) } \;\delta^2 r_+ \;\,.
\eea
We assume that the resulting spacetime is characterized by 
\be
r_+(\lambda) \;\,=\;\, r_+ \,+\, \lambda\, \delta r_+ \,+\, {\,\lambda^2 \over 2} \, \delta^2 r_+ \;,  \qquad\qquad
r_c(\lambda) \;\,=\;\, r_c \,+\, \lambda\, \delta r_c \,+\, {\,\lambda^2 \over 2} \, \delta^2 r_c \;\;,
\ee
the critical discriminant function $\,W_2(\lambda)\,$ up to second-order perturbation based on (\ref{crW2}) turns out to  be
\bea
W_2(\lambda) &=&  {1 \over 2} \left( -\, r_+ \,+\, \sqrt{\,1 \,-\, 2\, r_+^{2}\,} \right) \,\epsilon  \;\,+\;\, \frac{ \left( -\,1 \,+\, 4\, r_+^2 \right) \delta r_+ }{ -\, 1 \,+\, 2\, r_+^{2} \,+\, r_+ \sqrt{\,1 \,-\, 2\, r_+^{2}}  } \;\lambda \;\,-\;\, {1 \over 8} \left( -\, r_+ \,+\, \sqrt{\,1 \,-\, 2\, r_+^{2}\,} \right) \,\epsilon^2 \nn \\
&&~ +\;\, \frac{r_+ \,\delta r_+}{2 \left(-\,r_+ \,+\, \sqrt{\,1 \,-\, 2\, r_+^{2}}  \right) } \; \epsilon \,\lambda \;\,+\;\,  \left( \, {1 \over 2} \, \delta^2 r_+ \;\,+\;\, \frac{ \left(\,\delta r_+ \right)^2 \,+\, r_+ \, \delta^2 r_+ \,-\, 2 \, r_+^3 \,\delta^2 r_+ }{\left( \, 1 \,-\, 2\, r_+^{2} \, \right)^{3/2}}  \right.   \nn \\
&&~  +\; \left. \frac{\left( -\,1 \,+\, 2\, r_+ \sqrt{\,1 \,-\, 2\, r_+^{2}\,} \,\right) \left(\,\delta r_+ \right)^2 \;+\; r_+ \left( -\,1 \,+\, r_+^2 \,+\, 2\, r_+ \sqrt{\,1 \,-\, 2\, r_+^{2}\,}\,\right) \delta^2 r_+ }{2 \left(-\,3 r_+ \,+\, 5 r_+^3 \,+\, \left( \,1 \,+\, r_+^2 \,\right) \sqrt{\,1 \,-\, 2\, r_+^{2}\,} \, \right)} \,\right) \,\lambda^2 \;\,, 
\label{crW2IIv}
\eea
\end{widetext}
in which we have used (\ref{nex2}), (\ref{1varII}), and (\ref{2varII}). 
One can check easily that (\ref{crW2IIv}) is not a perfect square as those asymptotic flat black holes shown in \cite{Sorce:2017dst, Lin:2022ndf}. The value of $\, W_2(\lambda)\,$ is plotted in Fig. \ref{fig:2nd} in solid curves, showing clearly that the straight dashed lines due to constraint from linear order variation are modified to hyperbolas. 

We again choose $\,Q = 0.1\,$ which corresponds to extremal event horizon radius $\,r_+ = 0.102\,$. The blue solid curve in Fig. \ref{fig:2nd} corresponds to $\,\epsilon = 0\,$, showing again that the horizon for extremal RN-dS would definitely be destroyed by perturbation with $\,\lambda < 0\,$. As $\,\epsilon\,$ increases, however, the hyperbolas rise up and finally depart from the horizontal axis. The orange solid curve corresponds to a critical $W_2(\lambda)$ curve with $\epsilon=\epsilon_c:=0.119965$. This critical curve has its minimum occur at $W_2(\lambda)=0$ for some $\lambda$. Therefore, for $\epsilon > \epsilon_c$ with a given $Q$, the corresponding $W_2(\lambda)$ is positive definite so that WCCC holds. Compared with the orange dashed line obtained from linear order variation constraint, we see that the second-order effect makes it possible that $\, W_2(\lambda)\,$ become positive definite hence WCCC is restored. For $\,\epsilon > \epsilon_c\,$, e.g., the green curve with $\,\epsilon = 0.3\,$, the hyperbolas are all above the horizontal axis and $\,W_2(\lambda)\,$ are strictly positive definite. Our results show that the WCCC is violated only for near-extremal RN-dS. This is qualitatively consistent with the violation of SCC also for near-extremal RN-dS \cite{Cardoso:2017soq, Cardoso:2018nvb, Dias:2018ufh}.  However, the more precise comparison for the violations of SCC and WCCC cannot be done in principle due to the lack of relations between gravitational mass and matter mass in asymptotically de Sitter space.

\begin{figure}[hbt!]  
  \centering
  \includegraphics[width=1\linewidth]{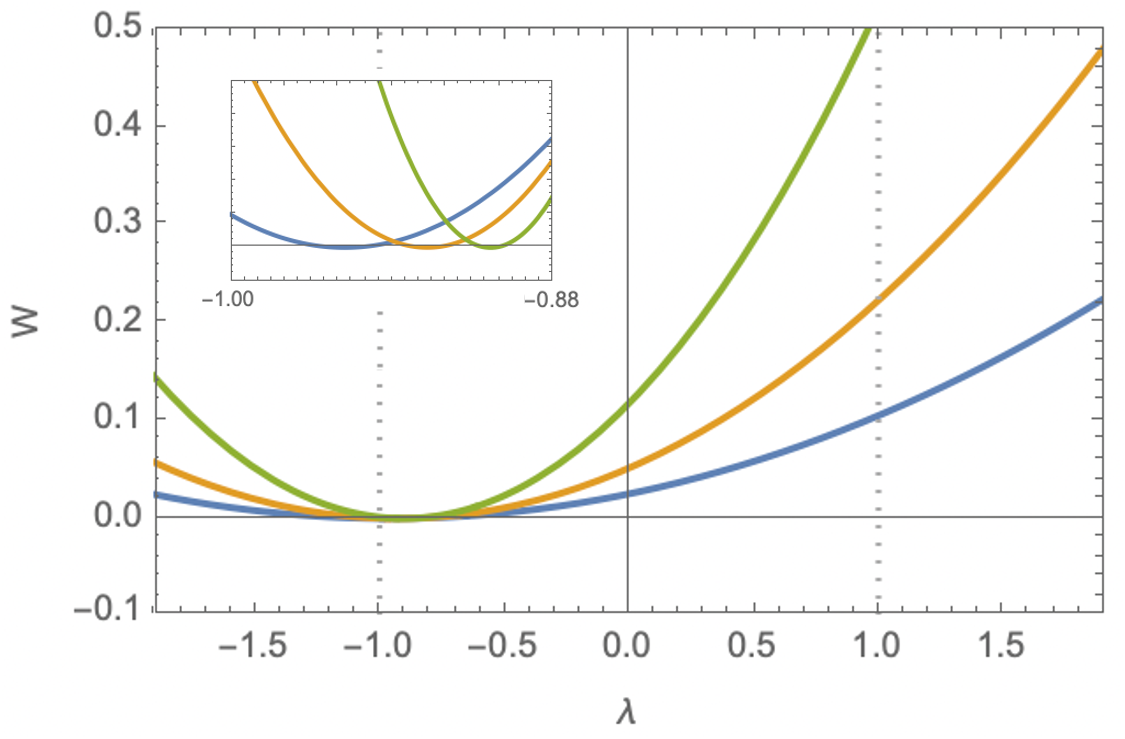}
  \caption{ The critical curves from the discriminant function $W_2(\lambda)\,$ saturating the first- and second-order second law with entropy given by \eq{S_W} for RN-dS of different values of $\, Q\,$. These critical curves are defined by having their local minimums occur at $W_2(\lambda)=0$ for some $\lambda$, and are thus characterized by the value of $Q$ and the corresponding value of $\epsilon_c$, which we denote the pair as $(Q,\epsilon_c)$. In this figure, $(Q,\epsilon_c)$ are $(0.05,  0.0539477 )$ for blue curve, $(0.1,0.119965)$ for orange curve, and $(0.2,  0.350888)$ for the green curve. We can also see that the minimums of these three curves all lie very closely and near $\lambda=-1$.} 
  \label{fig:Qc}
\end{figure}

Since our procedure is based on perturbation, which requires $|\lambda|<1$, we should trust the result when the perturbation approximation is valid. Therefore, we should be more careful with the above interpretation of our results to justify or falsify WCCC near the $\lambda=-1$ regime. Looking into this regime in Fig. \ref{fig:2nd} more closely, we see that the minimum of the orange critical curve seems rather near $\lambda = - 1\,$. We then like to see if the minimum of the critical $W_2(\lambda)$ curve can be shifted toward the smaller $|\lambda|$ regime by tuning the black hole parameter $Q$. The result is shown in Fig. \ref{fig:Qc} which shows the critical $W_2(\lambda)$ curves for different $Q$'s. These critical curves are defined by having their local minimums occur at $W_2(\lambda)=0$ for some $\lambda$, thus are characterized by the values of the pair $(Q,\epsilon_c)$.  We see that for quite a range of $Q$, the minimums of the critical curves are still near $\lambda=-1$. However, away from the regime of $\epsilon\approx \epsilon_c$\,, one can unambiguously identify the regime of violating WCCC for the (near-)extremal RN-dS inside the $|\lambda|\le 1$ window, as the blue line in Fig. \ref{fig:2nd}. Another interesting point, as shown in the inset in Fig. \ref{fig:Qc}, is that the minimum moves toward a slightly larger value of $\lambda$ (but smaller $|\lambda|$) as the value of $Q$ increases. However, $\epsilon_c$ increases as $Q$ increases. 


From the above discussions, it is fair to conclude that for such non-extremal black holes, small enough perturbation due to infalling matter could not destroy the black hole horizon. On the other hand, for very near extremal RN-dS, the second law constraint could not guarantee WCCC; the possibility of violating WCCC increases when closing to the extremal limit. This behavior is qualitatively in accord with the SCC result revealed in \cite{Cardoso:2017soq, Cardoso:2018nvb}.


%

\begin{figure}[hbt!]  
  \centering
  \includegraphics[width=1\linewidth]{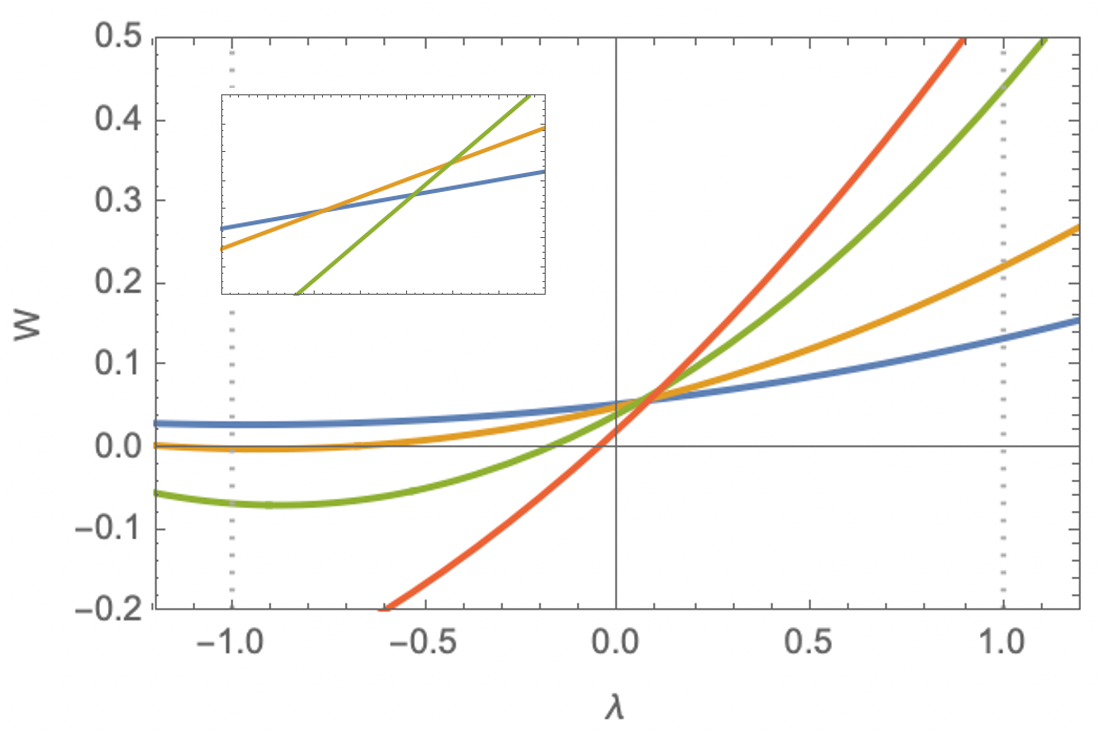}
 \caption{ The WCCC discriminant function $\, W_{2\,}(\lambda)\,$ saturating the first- and second-order second law with entropy given by \eq{S_W} for RN-dS of $\,\epsilon = 0.119965\,$ but for different $Q$ values, with $Q= 0.05$ (blue), $0.1$ (orange), $0.2$ (green) and $0.3$ (red), respectively. We also zoom in on the seemingly intersecting regime, as shown in the inset subfigure. The results imply that for a fixed sub-extremality, the WCCC tends to be violated for larger $Q$. }
  \label{fig:Qe}
\end{figure}

Moreover, we find that there exists no real $\epsilon_c$ when the parameter $Q$ is large enough, e.g., for $Q > 0.3$. This implies there exists no critical $W_2(\lambda)$ curve, and the WCCC must be violated in some regime of $|\lambda| \le 1$ for near-extremal RN-dS with $0< \epsilon\le 1$. Thus, it is interesting to see how the $W_2(\lambda)$ curve changes as we change $Q$ for a fixed $\epsilon$, and the results are shown in Fig. \ref{fig:Qe}. We can see that WCCC is easier to be violated for larger $Q$.

\section{Conclusion and Discussion}\label{sec_5}

In this paper, we have shown the regime of the parameter space of the near-extremal Reissner–Nordström black holes in de Sitter space (RN-dS), in which the weak cosmic censorship conjecture (WCCC) can be violated. The WCCC is examined based on Wald's gedanken experiment of overcharging the black hole by infalling matter, which causes the second-order variations obeying the second law constraint. Here, the second law constraint is to require the sum of the areas of event and cosmic horizons can never decrease. This is qualitatively comparable with the violation of strong cosmic censorship (SCC) for RN-dS configurations discovered earlier \cite{Cardoso:2017soq, Cardoso:2018nvb, Dias:2018ufh}.

The implication of our result is that the violations of WCCC and SCC could be correlated, even though they are formulated in quite different ways. Despite that, there are still some issues that remain to be explored further, and we will discuss them briefly before concluding our paper.

\begin{figure}[hbt!]  
  \centering
  \includegraphics[width=1\linewidth]{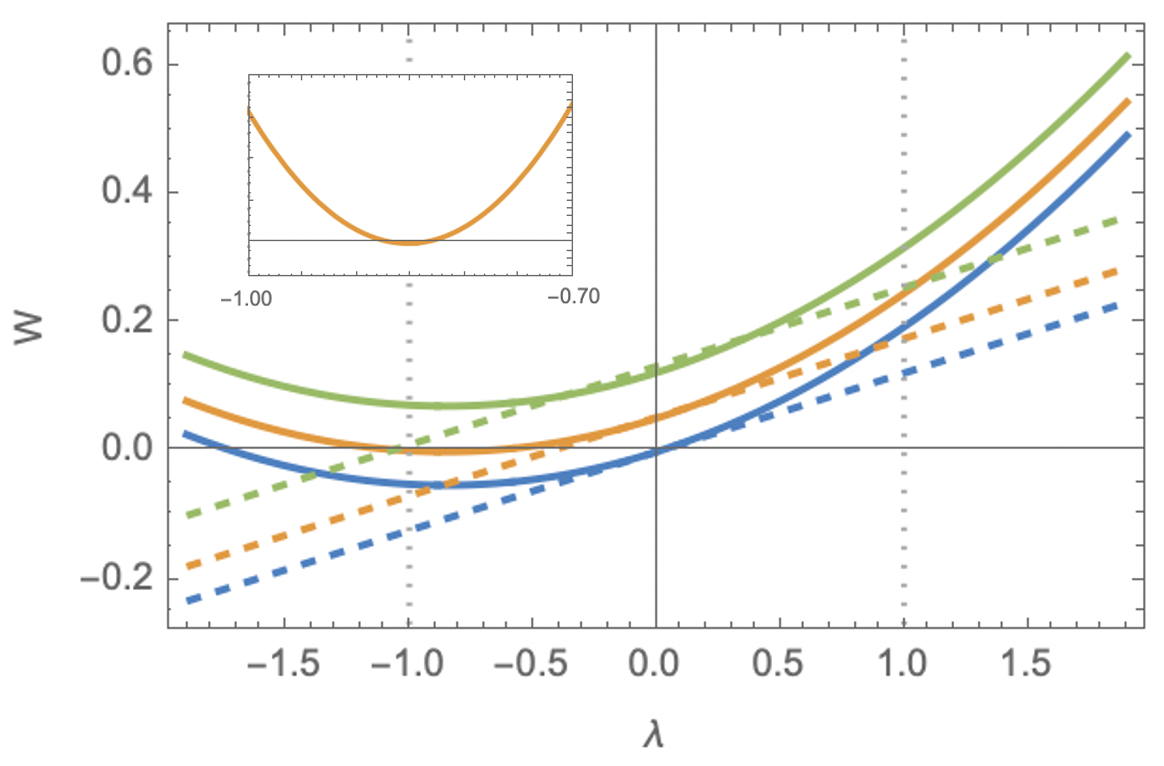}
  \caption{ The WCCC discriminant function $\, W_{2\,}(\lambda)\,$ saturating the first- and second-order second law with entropy given by $S=\pi r_c^2$. The results of corresponding $\,W_{1\,}(\lambda)\,$ are also shown in dashed lines. The line styles and parameters $Q$ and $\epsilon$ are chosen to be the same as in Fig. \ref{fig:2nd}. The results shown here look quite close to the ones in \ref{fig:2nd}. However, the critical value of $\epsilon$ is slightly changed to $\epsilon_c=0.121117$. } 
  \label{fig:2nd_rc}
\end{figure}

\begin{figure}[hbt!]  
  \centering
  \includegraphics[width=1\linewidth]{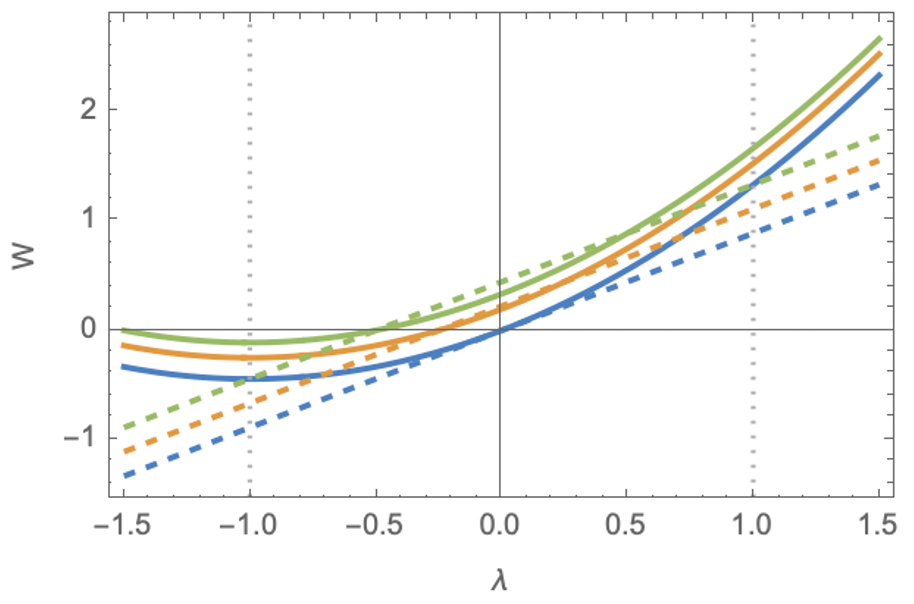}
  \caption{ The WCCC discriminant function $\, W_{2\,}(\lambda)\,$ saturating the first- and second-order second law with entropy given by $S=\pi r_+^2$. The results of corresponding $\,W_{1\,}(\lambda)\,$ are also shown in dashed lines. The line styles and parameter $Q$ are chosen to be the same as in Fig. \ref{fig:2nd} and \ref{fig:2nd_rc}, but here the blue, orange, and green lines correspond to $\epsilon=0$, $0.5$ and $1$, respectively. We see that the WCCC is violated even for $\epsilon 
  \ge 1$ cases. } 
  \label{fig:2nd_rp}
\end{figure}

For the eternal black hole in asymptotically flat spacetime, we can associate a Hawking temperature with the event horizon, and then the first and second laws of thermodynamics follow. Similarly, for pure de Sitter space. However, if there are both event and cosmic horizons, it seems that we cannot define a global temperature, and the whole spacetime may not be in thermal equilibrium to define the first law and, thus, the entropy for both the first and second laws. In the main text, we have adopted the definition of entropy to be the sum of the areas of the event and cosmic horizons (up to a constant factor). This is motivated by the meaning of entropy as the ignorance of the physical information shield behind both horizons. However, other choices for the entropy of the second law can also be considered to examine WCCC. For example,  it was argued based on the cosmic holographic entropy bound conjecture \cite{fischler1998holography, Bousso:1999xy} that the entropy of matter inside de Sitter space should be bounded by the order of the area of the cosmic horizon. If we adopt such entropy to examine WCCC based on our second-law method, we find that WCCC can be violated in a similar way as found in the main text with the chosen entropy \eq{S_W}, but with a slightly different regime of parameter space. The result is shown in Fig. \ref{fig:2nd_rc}. For curiosity, we also consider the case with entropy chosen to be the area of the event horizon, i.e., $S=\pi r_+^2$, and proceed to examine the WCCC and the result is shown in Fig. \ref{fig:2nd_rp}. Unlike the previous two cases, we see that WCCC is violated even for the $\epsilon=1$ case, which cannot be seen as the near-extremal RN-dS.
With the above comparison, we may say that the choices of entropy to be either $S=\pi (r_+^2 + r_c^2)$ or $S=\pi r_c^2$ are sensible because the corresponding regime for violating WCCC is qualitatively similar to the near-extremal regime for violating SCC.

The other issue that deserves future exploration is to try to formulate Wald's first-law approach to WCCC for RN-dS. As discussed in the Introduction, the difficulty of this approach is how to define the ADM mass to arrive at a relation like \eq{ADM}. Despite that, there are still some tentative proposals for defining gravitational mass in asymptotically de Sitter space \cite{Balasubramanian:2001nb, Dolan:2018hpl, Gao:2023luj}. We now consider the proposal of \cite{Dolan:2018hpl}, in which an Iyer-Wald invariant mass \cite{Iyer:1994ys} is defined on a spherical hypersurface at $r=r_{HS}$ with $r_+ <r_{HS} \le r_c$. This gravitational mass is equal to the mass parameter $M$ of RN-dS and is independent of the value of $r_{HS}$. Adopting this definition of gravitational mass, we can generalize the relation \eq{ADM} to RN-dS as $\delta M=E=-(m u_{\mu} + q A_{\mu}) t^{\mu}\big\vert_{r=r_{HS}}$, and yield a relation similar to \eq{1st_law} as follows,
\be \label{WCCC_1st_law}
\delta M -  {Q\over r_+} \delta Q \ge 0
\ee
where we have used $A_{\mu}t^{\mu}\big\vert_{r=r_+}=-{Q\over r_+}$ with $r_+$ given by \eq{rp} for an extremal RN-dS, and also $\delta Q=q$ by Gauss's law. On the other hand, the extremal condition \eq{ex} for RN-dS can be rewritten into (set $L=1$)
\be\label{ex_1}
54 M^2 = \Big(2+\sqrt{1-12 Q^2}\Big)^2 \Big(1-\sqrt{1-12 Q^2}\Big)\;.
\ee
The sub-extremal condition for RN-dS is obtained by replacing the $\,=\,$ in \eq{ex_1} by $\,\ge\,$. Furthermore, by replacing $M$ by $M+\delta M$ and $Q$ by $Q+\delta Q$ in this sub-extremal condition and expanding up to the first order of and $\delta M$ and $\delta Q$, we can obtain the condition for WCCC of an extremal RN-dS, which turns out to be nothing but \eq{WCCC_1st_law}. This implies that WCCC is preserved for the extremal RN-dS by adopting the first law relation \eq{WCCC_1st_law} with $\delta M$ defined in the way of \cite{Dolan:2018hpl}.

The above WCCC-preserving result seems in contradiction with our WCCC-violating result as shown in Fig. \ref{fig:1st} obtained from the second-law method. This implies that one of the two methods should be flawed. In the second-law approach, we can understand that the violation of WCCC mainly comes from the increase in the area of the cosmic horizon, which compensates for the shrinking of the event horizon. However, in the relation \eq{WCCC_1st_law}, there is no term reflecting the area change of the cosmic horizon. Thus, it is more like a local statement associated with the event horizon only.  On the other hand, if we think the gravitational mass can be defined quasi-locally in the way of \cite{Dolan:2018hpl}, then once the particle enters the black hole, the mass and charge parameters in RN-dS will also change accordingly, so does the radius of $r_c$ for a given $L$. Indeed, our second-law approach does take such change into account. The above arguments imply that any first law for RN-dS without taking care of the change of entropy associated with the cosmic horizon, such as the one given by \eq{WCCC_1st_law}, will be incomplete and flawed. The flaw of the first-law approach could be due to the quasi-local definition of the gravitational mass of RN-dS in the way of \cite{Dolan:2018hpl}. However, a good definition of gravitational mass in asymptotically de Sitter to take the cosmic dynamics into account is still out of reach and deserves further study. 

\begin{acknowledgments}
We thank Jason Payne for the discussion on WCCC of Myer-Perry's BHs.
FLL is supported by Taiwan's National Science and Technology Council through Grant No.~112-2112-M-003 -006 -MY3.  BN is supported by the National Natural Science Foundation of China with Grant No.~11975158 and No.~12247103.
\end{acknowledgments}

\appendix

\section{WCCC of 5D Myers-Perry Black Holes}\label{app_1}

In this appendix, we examine the WCCC for the 5D Myers-Perry black holes \cite{Myers:1986un}. The metric in Boyer-Lindquist coordinates is 
\bea
ds^2 &=& - \,dt^2 \,+\, \frac{\rho^2}{4\Delta}dx^2 \,+\, \rho^2 d\theta^2  \nn \\
&& +\,\, (x + a^2) \sin^2{\theta} \,d\phi^2  \,+\,  (x + b^2) \cos^2{\theta} \, d\psi^2  \nn \\
&& +\,\, \frac{r_0^2}{\rho^2} \left( \,dt \,+\, a \sin^2 \,\theta \,d\phi \,+\, b \cos^2 \theta \, d\psi  \right)^2 ~,
\eea
in which 
\bea
\rho^2 &=& x \,+\, a^2 \cos^2 \theta \,+\, b^2 \sin^2 \theta \;,  \nn \\
\Delta &=& (x + a^2)(x + b^2) \,-\, r_0^2 x \;. \nn 
\eea
The parameter $\,r_0\,$ is related to the mass $\,M\,$ of the Myers-Perry black hole, 
\be
M ~=~ { 3 \pi \over 8 G} \,r_0^2 \;, 
\ee
and $\,a, \,b\,$ is related to the two independent angular momentums $\,J_a,\,J_b\,$:
\be
J_a ~=~ { 2 \over 3 } M a \;,  \qquad
J_b ~=~ { 2 \over 3 } M b \;.
\ee
The locations of the horizons are indicated by $\,\Delta = 0\,$, which gives 
\be
x_\pm ~=~ {1 \over 2} \left( \, r_0^2 \,-\, a^2 \,-\, b^2 \,+ 
\sqrt{ \,( r_0^2 \,-\, a^2 \,-\, b^2 )^2 \,-\, 4 a^2 b^2 } \, \right) \;.
\ee
The angular velocities on the event horizon are 
\be
\Omega_a ~=~ { a \over x_+ \,+\, a^2 } \;,  \qquad
\Omega_b ~=~ { b \over x_+ \,+\, b^2 } \;.
\ee
If $\, r_0 \,=\, a \,+\, b \,$, the black hole is extremal with $\,x_+ \,=\, x_-\,$, hence we define the following sub-extremal condition
\be
W(r_0, \,a, \,b) \;\,=\;\, r_0^2 \,-\, (a \,+\, b)^2 \;,
\ee
and introduce the sub-extremal parameter $\,\epsilon\,$ by
\be
r_0 \;\,=\;\, \sqrt{\,(a \,+\, b)^2 \,+\, \epsilon^2 } \;.
\ee
The black hole entropy is 
\be
S \;\,=\;\, { \pi^2 r_0^2  \over 4 \,G \,\kappa} \, \left( \, 2 \,-\, \frac{ 2 \,a^2 }{x_+ +\, a^2} \,-\, \frac{ 2\, b^2 }{x_+ +\, b^2} \,\right) \;,
\ee
with $\,\kappa\,$ the surface gravity 
\be
\kappa \;\,=\;\, { 2 x_+ \,+\, a^2 \,+\, b^2 \,-\, r_0^2 \over r_0^2 \sqrt{\,x_+} }  \;.
\ee
The Hawking temperature is simply $\,T_{\rm H} \,=\, \kappa / (2\pi)\,$. First, we consider linear order perturbation. The linear order variation inequality $\,\delta^1 S(r_0, \,a,\,b) \ge 0\, $ gives 
\be \label{linear1MP}
  \delta r_0 \;\,=\;\, \delta a \,+\, \delta b \,-\,
\frac{ b\,( 5 a \,+\,b) \,\delta a \,+\, a \,(5 b \,+\, a) \,\delta b }{\sqrt{a b} \,(a \,+\, b)^2} \,\epsilon   ~, 
\ee
Assuming the perturbed metric finally settles down to another Myers-Perry black hole characterized by 
\bea
r_0\,(\lambda) &=& r_0 \,+\, \lambda\, \delta r_0  \;,  \nn \\
a\,(\lambda) &=& a \,+\, \lambda\, \delta a  \;,   \\
b\,(\lambda) &=& b \,+\, \lambda\, \delta b  \;,  \nn 
\eea
It turns out that the critical discriminant function $\,W_1(\lambda)\,$ is null 
\be\label{W1MP}
W_1(\lambda) \;=\; 0 \;,
\ee
that is, the WCCC is always preserved with respect to linear order perturbation. To see the effect of the second order perturbation, we set the linear order perturbation to be optimally done, i.e., $\,\delta^1 S = 0\, $, up to $\,{\cal O}(\epsilon^2)\,$ giving  
\begin{widetext}
\be \label{linear2MP}
  \delta r_0 \;\,=\;\, \delta a \,+\, \delta b \,-\,
\frac{ b\,( 5 a \,+\,b) \,\delta a \,+\, a \,(5 b \,+\, a) \,\delta b }{\sqrt{a b} \,(a \,+\, b)^2} \,\epsilon  
\, + \, \frac{ b^2 \, ( \,41 a^2 \,+\, 10 a b \,+\, b^2 \,) \,\delta a \,+\, a^2 \, ( \, 41 b^2 \,+\, 10 a b \,+\, a^2 \,)\, \delta b  }{2 a b \,(a \,+\, b)^4}\,\epsilon^2 ~, 
\ee
\end{widetext}
while the second order variation inequality $\,\delta^2 S \ge 0\, $ gives
\bea \label{secondMP}
\delta^2 r_0 &\ge& { 1 \over a b \,(a \,+\, b)^3} \, \left( \, b^2\, ( \,25 a^2 \,+\, 10 a b \,+\, b^2 \,) \,(\delta a)^2 \right. \nn \\
&&  + \;\, a^2 \,( \,25 \, b^2 \,+\, 10 a b \,+\, a^2 \,) \, (\delta b)^2   \nn \\
&&  + \;\,  a b \left.  (\, 10 a^2 \,+\, 52 a b \,+\, 10 b^2 \,) \,\delta a \delta b   \,\right)  \nn \\
&&  + \;\, \delta^2 a \,+\, \delta^2 b  \,+\, {\cal O}(\epsilon)\;. 
\eea
Assuming that the resulting metric is described by 
\bea
r_0(\lambda) &=& r_0 \,+\, \lambda\, \delta r_0 \,+\, {\,\lambda^2 \over 2} \, \delta^2 r_0 \;,  \nn \\
a(\lambda) &=& a \,+\, \lambda\, \delta a \,+\, {\,\lambda^2 \over 2} \, \delta^2 a \;,   \\
b(\lambda) &=& b \,+\, \lambda\, \delta b \,+\, {\,\lambda^2 \over 2} \, \delta^2 b \;\;,  \nn 
\eea
the critical discriminant function $\,W_2(\lambda)\,$ turns out to be a perfect square
\be \label{W2MP}
W_2(\lambda) \;=\; \left( \,\epsilon \;-\; \frac{ \left(\, b^2 \delta a \,+\, a^2 \delta b + 5 a b \,(\,\delta a \,+\, \delta b\,)\, \right)}{\sqrt{a b } \;(a\,+\,b)} \,\lambda \,\right)^2 \;\,,
\ee
showing that the WCCC is also satisfied up to the second-order perturbation.  

The above result is also consistent with the Sorce-Wald procedure \cite{Sorce:2017dst} based on the first law. The linear order variation inequality (assuming null energy condition) is
\be \label{SW1stMP}
\delta M \,-\, \Omega_a \delta J_a \,-\,\Omega_b \delta J_b \;\ge\; 0  \;,
\ee
in which
\bea
\delta M &=&  { d M \over d r_0 }\, \delta r_0 ~,   \\
\delta J_a &=& { \partial J_a \over \partial r_0 } \, \delta r_0 \;+\; {
\partial J_a \over \partial a } \,\delta a ~,  \\
\delta J_b &=& { \partial J_b \over \partial r_0  } \, \delta r_0 \;+\; {
\partial J_b \over \partial b } \,\delta b  \;.    
\eea
Up to $\,{\cal O} (\epsilon)\,$, we find (\ref{SW1stMP}) simply gives (\ref{linear1MP}), hence leads to (\ref{W1MP}). Assuming that the linear order perturbation is optimally done, up to $\,{\cal O} (\epsilon^2)\,$ we obtain (\ref{linear2MP}). The constraint for the second-order perturbation should be 
\be \label{SW2ndMP}
\delta^2 M \,-\, \Omega_a \,\delta^2 J_a \,-\, \Omega_b \,\delta^2 J_b 
\;\ge\; - \,T_{\rm H}\, \delta^2 S^{\rm MP}   \;,
\ee
in which the second-order change in mass and angular momentums are given by 
\bea
\delta^2 M &=&  { d M \over d r_0 }\, \delta^2 r_0 ~,   \\
\delta^2 J_a &=& { \partial J_a \over \partial r_0 } \, \delta^2 r_0 \;+\; {
\partial J_a \over \partial a } \,\delta^2 a ~,  \\
\delta^2 J_b &=& { \partial J_b \over \partial r_0  } \, \delta^2 r_0 \;+\; {
\partial J_b \over \partial b } \,\delta^2 b  \;,    
\eea
while $\,\delta^2 S^{\rm MP}\,$ denotes the second-order change in the entropy for a one-parameter family of Myers-Perry black holes with parameters given by
\bea
r_0^{\rm \,MP} &=& r_0 \,+\, \alpha \,\delta r_0  \;,  \\
a^{\rm \,MP} &=& a \,+\, \alpha \,\delta a  \;,   \\
b^{\rm \,MP} &=& b \,+\, \alpha \,\delta b  \;. 
\eea
Thus,
\bea
\delta S^{\rm MP} &=& {\partial S \over \partial r_0} \,\delta r_0 \,+\, 
{\partial S \over \partial a } \,\delta a \,+\, {\partial S \over \partial b}\, \delta b \;,   \\
\delta^2 S^{\rm MP} &=& {\partial \left(\delta S^{\rm MP} \right) \over \partial r_0} \,\delta r_0 \,+\, 
{\partial \left(\delta S^{\rm MP}\right) \over \partial a } \,\delta a \,+\, {\partial \left(\delta S^{\rm MP}\right) \over \partial b}\, \delta b \,.   \nn \\ ~
\eea
We find that (\ref{SW2ndMP}) finally gives rise to exactly (\ref{secondMP}), hence leading to the same perfect square (\ref{W2MP}).

\bibliography{ref.bib}

\end{document}